\documentclass[thmsa,titlepage,draft,12pt]{article}
\usepackage{sw20lart}

\input tcilatex
\QQQ{Description}{
This is a paper including both 'the special theory of quantum time' and
'the general theory time'. Networks of Quantum clocks are the basis of
macroscopic time. 
}

\QQQ{Typist}{
SMH
}

\QQQ{Language}{
American English
}

\begin{document}

\author{Scott Hitchcock \\
National Superconducting Cyclotron Laboratory\\
Michigan State University, East Lansing, MI 48824-1321\\
NSCL Publication: MSUCL-1123\\
E-mail: hitchcock@nscl.msu.edu}
\title{\textbf{Quantum Clocks and the Origin of Time in Complex Systems.}}
\date{February 20, 1999}
\maketitle

\begin{abstract}
The origin and nature of time in \emph{complex} \emph{systems} is described
with a \emph{special} and a \emph{general theory of time}. The special
theory of time describes \emph{quantum clocks }and their role as sources of 
\emph{signals }(e.g. decay or collision products). The method of Feynman
diagrams is used to define a \emph{Feynman clock}. The general theory of
time describes \emph{networks} of quantum or Feynman clocks and their
signals. These networks form the basis of evolution in complex systems.

A\emph{\ temporal correspondence principle} describes \emph{temporal} \emph{%
phase} \emph{transitions} from \emph{collective excitation }to \emph{%
discrete n-sum representations }of quantum clocks in a network. \emph{%
Temporal phase transitions} mark the emergence of classical properties of
systems such as \emph{irreversibility}, \emph{entropy}, and \emph{%
thermodynamic arrows of time} in the evolution of the universe.

The general concept of \emph{time} is translated into the \emph{lifetimes}
of \emph{unstable} configurations of matter. The application of the special
and general theories of time to quantum cosmology are discussed.(Revised
February 20, 1999).

$\QTR{sl}{Keywords}$\emph{:} The Problem of Time, the Arrow of Time, Time
asymmetry, Entropy and Time, Collective Excitations, Time Measurement,
Atomic Clocks, Many-Body Problem, Complexity and Nonlinear Systems, Fractal
Time, Cellular Networks, the Wheeler-DeWitt Equation, Instantons, and
Quantum Cosmology.
\end{abstract}

\tableofcontents

\section{Introduction}

Time is one of the most important features of our changing world. The cause
and effect relationship between events is thought to be guided by
irreversible 'arrows of time' pointing from 'what was' to 'what is' and then
to 'what will be'. These irreversible changes define the course of our lives
and the evolution in the world around us. The \emph{cause} of this change
and the time associated with it is one of the least understood aspects of
our universe. In spite of this lack of understanding about the origin of
time we have developed 'quantum' or atomic clocks to a very high degree of
accuracy \cite{Ashworth}, \cite{Atomic clock}. We use them to 'measure'
change, control our activities and define our perceptions of reality.

Time is often viewed as a 'parameter' in the description of dynamic quantum
systems \cite{Sakurai}. As a parameter 'time' does not have the same
properties as other quantum observables. In general relativity time is
converted into a 'spatial' measure of the distance between events. This
spatialization of time arises when two distinct causally connected events
are mapped onto separate hypersurfaces where the '\emph{time}' between the
events is transformed into a '\emph{distance}' (\emph{lapse function})
interval \cite{Reich}, \cite{Kolb}. The \emph{time} created in this way
implicitly assumes the result (directionality of time) that is to be proved
(''petitio principii''). This paradox can be resolved by \emph{defining}
time as a \emph{lifetime.}

The quantum world is the \emph{source} of all the irreversible changes we
observe. The \emph{special theory of time} is based on the finite \emph{%
lifetimes} of \emph{unstable} quantum systems or \emph{quantum clocks}.
Quantum clocks may be created from the decay of a particle or the
collisional interactions of particles in a volume of space $V_{qc}$. The
lifetimes of these systems can be calculated using the methods of \emph{%
Feynman diagrams}. These systems will be called \emph{Feynman clocks}.

'Unstable' applies to any \emph{isolated }system (quantum or classical) in
an excited, perturbed or non-equilibrium state. The decay of an unstable
system is \emph{caused} by the asymmetric distribution of its masses and
charges under the influence of repulsive and attractive fundamental
interactions (e.g. strong, electromagnetic, weak, gravitational forces or
their combinations at high energies in the early universe). An unstable 
\emph{discrete} initial state decays to another discrete reconfiguration
state from within a \emph{continuum} of possible states \cite{Diu}. This new
state of the system may subsequently decay as part of a chain reaction.
These chain reactions form the basis of \emph{networks}. Networks map the
evolution of complex systems as sequences of causes and effects. A set of 
\emph{axioms} define the properties of Feynman clocks. In modified \emph{%
cellular networks} \cite{cellnets} the vertices are Feynman clocks\emph{\ }%
and the arcs are signal trajectories. With the following substitutions,
Quantum or Feynman clocks for nodes and Signals for bonds (along with
modifications of the assumptions of the space-time picture; see axioms), the
Cellular networks of M. Requardt \cite{cellnets2} provide an excellent
mathematical framework for the exploration of the evolution of complex
systems. These \emph{Feynman networks} are used to develop a \emph{general
theory of time} that describes the \emph{emergence} of \emph{macroscopic
lifetimes }in\emph{\ complex systems}. Once the axioms are applied, the
tools already developed by others for dealing with these networks can be
applied to model \emph{message} (i.e. information content of network
signals) transmission in many body networks \cite{Messages}.

The decay process may also result in shape or positional changes of the
reconfigured system with resulting signals in the form of \emph{images.} The
image can be the combination of \emph{emitted } and \emph{scattered signals}.

Reconfigurations of macroscopic systems can be detected as changes or
transitions in \emph{thermodynamic} coordinates such as temperature,
pressure and volume changes, etc. The lifetime of the states associated with
these thermodynamic transitions is determined from the network of $n$
quantum clocks acting together as either a single \emph{collective quantum
excitation} or a classical sequential sum ($n$\emph{-sum}) of clock and
signal lifetimes. The \emph{temporal correspondence principle} determines
which description is needed based on the magnitude of the '\emph{action}'
for the system and the strength of the \emph{coupling} between the quantum
clocks. A\emph{\ temporal phase transition }occurs when a complex
interactive system in a collective excitation state \emph{decouples} into a
set of distinct uncoupled quantum clocks. The transition may be created by
increased spatial separation of components due to a change in the potential
energy between clocks. At a \emph{critical distance} the clocks then
decouple. A \emph{critical energy} for system may be reached by energy
losses translating into decreased interaction strengths between clocks. The
decay of a phonon state resulting in system energy loss through the
scattering of a signal can return a system to a decoupled state.

The largest example of a quantum clock may be the universe. Its 'decay' from
an unstable state of extreme high energy is the 'first cause' for the
creation decay products such as matter, space (the vacuum), and the complex
evolving systems we observe today (see Section 5).

\section{Quantum Clocks}

The \textbf{special and general theories of time} based on \textbf{quantum
clocks} in networks begins with the following axioms:

\begin{axiom}
A \textbf{clock}\emph{\ }is an\emph{\ }\textbf{unstable system} or \textbf{%
source} that decays with a \textbf{finite lifetime}.
\end{axiom}

\begin{axiom}
A \textbf{quantum clock} is a system that requires a quantum mechanical
description. This occurs when the '\textbf{action}' (energy $\times $ time)
of the system is of the order of \textbf{Planck's constant} ($\hbar $) \cite
{White}.
\end{axiom}

\begin{axiom}
\textbf{Clocks} emit or produce \textbf{signals} when they \textbf{decay}. 
\textbf{Signals}\emph{\ }can be \textbf{detected} by absorption in another
system which may become unstable resulting in the creation of another
signal. An unstable system can also become stable with no further signal
creation.
\end{axiom}

\begin{axiom}
Two types of time are associated with quantum clocks. \textbf{Intrinsic time}
is the sum of the \textbf{decay lifetime}\emph{\ }of an excited state of a 
\textbf{clock} and the \textbf{initialization lifetime}\emph{\ }of the
preparation or reconfiguration process that sets up the excited state of the
clock. The initialization lifetime is the lifetime of the \textbf{%
reconfiguration message} created in the detector by its carrier signal.
Note: decay lifetime is used interchangeably with intrinsic lifetime
henceforth unless otherwise noted. The other \textbf{time} is the \textbf{%
signal lifetime. }This is the lifetime starting with the creation of the
signal by the source and ending with its absorption at a detector. This is
equal to the classical translation time of an object or signal in space.
\end{axiom}

\begin{axiom}
Sources may be created by collisions of particles or the merging of two or
more systems. The decay products connect emission sources to detection sites
with signals that form a \textbf{network}.
\end{axiom}

\begin{axiom}
Clocks may be \textbf{reset} to an unstable configuration by\emph{\ }\textbf{%
detection}\emph{\ }of \textbf{signals}. \textbf{Feedback} can reset a system
many times exhibiting cyclical \textbf{self-regulation}. Regular repeated
detection of signals is the basis for a \textbf{standard} \textbf{clock}.
The number of reset-decay cycles for a clock can range from 1 to infinity. A 
\textbf{zero order clock} is a system that decays once and is not reset
(e.g. the initial state of the universe).
\end{axiom}

\begin{axiom}
There are two '\textbf{arrows of time}'. The \textbf{intrinsic arrow} always
'points' or maps energy changes causing the decay of \textbf{unstable states}
to lower energy \textbf{decay states}. This is the lifetime associated with
network \textbf{vertices}. The \textbf{signal arrow} always 'points' from
the \textbf{source}\emph{\ }\textbf{system} to a \textbf{detector}\emph{\ }%
\textbf{system} (or \textbf{sink} in the case of empty space). This is the
lifetime associated with the signals or \textbf{arcs}\emph{\ }connecting
vertices in network.
\end{axiom}

\begin{axiom}
For networks with $n$ vertices and $n$ arcs, the total lifetime or \textbf{%
n-sum lifetime} is a \textbf{classical} \textbf{sum} of the\textbf{\ }%
lifetimes for the individual clocks and signals acting \textbf{sequentially}.
\end{axiom}

\begin{axiom}
The '\textbf{collective excitation lifetime}' is a \textbf{quantum}
description \cite{Mattuck} of a system of $n$ quantum clocks acting \textbf{%
collectively}\emph{\ }\textbf{as a single unstable quantum system} (e.g.
phonons, plasmons, etc.).
\end{axiom}

\begin{axiom}
The \textbf{n-sum }and \textbf{collective excitation}\emph{\ }descriptions
make a \textbf{temporal phase transition }when the signal path distances or 
\textbf{bond} energies between quantum clocks reach critical values. At this
scale the \textbf{temporal correspondence principle}\emph{\ }defines the
transition boundary conditions for \textbf{decoupling} of a \textbf{%
collective}\emph{\ }\textbf{excitation quantum clock} into an \textbf{n-sum
set of quantum clocks}. Isolated and collective quantum clocks along with
their signals can create complex macroscopic networks with classical
temporal features.
\end{axiom}

It is important to note that in general, interpretation of the information
content in signals involves comparisons with \textbf{standard clocks}. The
lifetimes of unstable signal generating systems are 'measured' through a
process of \textbf{signal mapping} to a standard (quantum) clock. Note that $%
\hbar =c=1$ unless they are expressly shown in the following equations.

\section{The Special Theory of Time}

The quantum clock is the building block of all macroscopic and large scale
structures in the observable universe. A simple quantum clock is described
using \textbf{time independent perturbation}\emph{\ }\textbf{theory, Fermi's
Golden Rule,} and a \textbf{time-independent Hamiltonian} transforming a 
\textbf{discrete}\emph{\ }\textbf{resonantly coupled (excited) state} to
another state from within a \textbf{continuum }of the \textbf{possible
reconfiguration (lower energy) states} after decay. The initial discrete
state decays \textbf{irreversibly} with a \textbf{finite}\emph{\ }\textbf{%
lifetime}. This is the point at which directional causality appears.

Two important points are made here. First a \textbf{time independent process}
creates a '\textbf{time}' which for a quantum clock is the \textbf{lifetime}
of that process. The key to this \textbf{dimensional conversion} of the 
\textbf{energy-momentum reconfiguration }of a quantum clock to a \textbf{%
temporal representation} is \textbf{Planck's constant}, $\hbar $. This
dimensional conversion represents the subtle connection between temporal
irreversibility of complex reconfiguration processes and the interactions of
energy and matter.

The second point is that the \textbf{continuous nature} of the set of
possible final reconfiguration states \textbf{causes} the \textbf{%
irreversible decay.} The \textbf{intrinsic arrow of time} is directly the
result of this resonant coupling of the initial state to the continuum. The%
\textbf{\ finite lifetime} of the initial state is \emph{caused} by this
coupling to another state of the\textbf{\ same energy} in the continuum set.
This coupling to states of \textbf{different energy} causes the \textbf{%
energy shift},\textbf{\ }$\delta E$ , in the initial discrete state energy.
This shift defines the location of the center ($E_{discrete}+\delta E$) of
the \emph{Lorentzian energy distribution} of the final states after decay.
This distribution maps the dispersion in the energy of final states from
which the probability for the system to be in a given reconfiguration state
is determined. The \emph{full width at half maximum} of this distribution is 
$\Gamma $. The \textbf{lifetime} of this process is the \emph{inverse} of
the width of the final Lorentzian energy distribution (see Complement D$_{%
\text{XIII}}$ in \cite{Diu} for details of this description): 
\begin{equation}
\tau =\frac \hbar \Gamma 
\end{equation}

This simple decay lifetime is generalized for the many-body problem by
looking at the \emph{momenta} of the components of quantum clocks. In this
case a decay process begins with the initial momentum, $\mathbf{p}_0$, of
the discrete unstable state. The general equation for the intrinsic lifetime 
\cite{Veltman} of this isolated quantum clock is given by:

\begin{eqnarray}
\mathbf{\tau }_{qc} &=&\dfrac \hbar {\mathbf{\Gamma }_I} \\
&=&\tfrac \hbar {\didotsint\limits_{q_n\cdots q_1}\left[ \tfrac{V^{n+1}}{%
\left( 2\pi \right) ^{3n+4}}\mathbf{P\cdot }\left| \mathbf{M}_I\right|
^2\delta _4\left( \mathbf{p}_0-\tsum\limits_{j=1}^n\mathbf{q}_j\right)
\right] dq_1dq_2\cdots dq_n^{}}
\end{eqnarray}
where $\mathbf{\Gamma }_I$ is the decay rate (in $MeV$). $V$ is the 3-space
volume of the quantum clock. The permutational factor $\mathbf{P}$ keeps
track of the number of identical particles produced by the process. The
reduced matrix element, $\mathbf{M}_I$ is equal to the usual $S$-matrix
element, apart from the $\delta $-function for overall energy-momentum
conservation. The index $I=s,em,w,$or $g$ represents the strong,
electromagnetic, weak, or gravitational forces while including the
possibility of \emph{coupled} fundamental interactions (e.g. the electroweak
or GUTs epoch unified interaction in the case of high energy states in the
early universe). The $\mathbf{M}_I$ matrix contains all the physical
information for the decay process including the way in which the fundamental
interactions drive the reconfiguration. If the reconfiguration term: 
\begin{equation}
\mathbf{M}_I=0
\end{equation}
then the 'lifetime' of the transition is \emph{infinite }(i.e. no decay
occurs). The initial momentum for the unstable system is $\mathbf{p}_0$. The
momenta of the decay products (\emph{signals}) are $\mathbf{q}_j$.

The lifetime, $\tau _{qc}$, of a decay transition between configurations
represents the intrinsic 'tick' (lifetime) of the clock. This 'tick' is
transformed into state information carried by the \emph{signal} produced in
the decay process. An example of this process is the creation of the
universe from the irreversible decay of an unstable initial \emph{%
non-stationary} state due to an 'initiation' perturbation (quantum
fluctuation) in the total energy (see Section 5).

A general description for the\textbf{\ intrinsic lifetime} of an unstable
system produced by decay or \textbf{collisions} (a \textbf{Feynman clock})
is the same as above except that the initial state with momentum $\mathbf{p}%
_0$ is a composite discrete unstable state \emph{created} in the space of
volume $V$ by the interactions (collisions) of the incoming momenta $\mathbf{%
p}_i$. With the substitution

\begin{equation}
\mathbf{p}_0=\tsum\limits_{i=1}^m\mathbf{p}_i
\end{equation}

we have the \textbf{lifetime }for a\textbf{\ Feynman Clock} given by:

\begin{eqnarray}
\mathbf{\tau }_{fc} &=&\dfrac \hbar {\mathbf{\Gamma }_I} \\
&=&\tfrac \hbar {\didotsint\limits_{q_n\cdots q_1}\left[ \tfrac{V^{n+1}}{%
\left( 2\pi \right) ^{3n+4}}\mathbf{P\cdot }\left| \mathbf{M}_I\right|
^2\delta _4\left( \tsum\limits_{i=1}^m\mathbf{p}_i-\tsum\limits_{j=1}^n%
\mathbf{q}_j\right) \right] dq_1dq_2\cdots dq_n}
\end{eqnarray}

where the $\mathbf{q}_i$ are the signals (or scattered momenta). When using
this method a Feynman clock is defined by the incoming and outgoing momenta
in a specified volume $V$ with \emph{overall} \emph{energy-momentum} \emph{%
conservation}.

The outgoing particles of momenta $\mathbf{q}_j$ (signals) created by a
Feynman clock have a \textbf{signal lifetime}\emph{\ }given by the distance, 
$d_{q_j}$ (geodesic path length calculated with an appropriate metric for
the space) traveled from the \textbf{source} to a \textbf{detector} divided
by the propagation velocity, $v_{q_j}$:

\begin{equation}
\mathbf{\tau }_s=\frac{d_{q_j}}{v_{q_j}}
\end{equation}

\subsection{Signal Paths Between Events}

The incoming momenta in a Feynman clock are time-independent in an \emph{%
operational or empirical} description of a system. An example of this is a
spectrogram. Energies and momenta of photons are \emph{determined} from the
spectral distribution function created by interpretation of events recorded
by a detection system. The \emph{information} created after the \emph{%
detection} of a set of spectral signals is causally and temporally
independent of their sources.

For this reason the decay or signal momenta are treated as 3-momenta in a
3+0 space (without 'time'). The Minkowski 4-metric for a 3+1 spacetime
geometry transforms from 
\begin{equation}
ds_4^2=c^2dt^2-dx_1^2-dx_2^2-dx_3^2
\end{equation}
to a 3-metric for the space alone: 
\begin{equation}
ds_3^2=dx_1^2+dx_2^2+dx_3^2
\end{equation}
since the \emph{spatialized time term} ($c^2dt^2$) becomes $0$ when
transforming events into a \emph{3-space} reference frame using the \emph{%
lifetimes} of quantum clocks and their signals to map the evolution and
reconfigurations of systems. This assumption applies to any metric that
couples time to space. Relativistic corrections can be applied to these
systems when the signal and decay 'lifetimes' are substituted for the
relativistic or proper 'time' between \textbf{events}. The \emph{locations}
of the signal emission sources and signal detection systems in this
time-independent 3-space are the \emph{events} of conventional \emph{%
spacetime}. The signals from these unstable systems provide the basis of
causal connectivity between events (source and detector quantum clocks).

The intrinsic and signal lifetimes are real numbers associated with the
vertices (events) and arcs (signals) of a causal network. By transforming
the spatialized time term into a lifetime we can prepare the foundation for
developing a ''\emph{background-free}'' \cite{Baez} theory of time. Time is
neither a coordinate or dimension in this case but the magnitude of an
intrinsic or signal lifetime originating from an unstable system.

\subsection{Virtual Quantum Clocks and Signals.}

A special case arises for a spatial scale (of the order of the Planck length
or about $1.6\times 10^{-35}$ meters \cite{Baez}, \cite{cellnets}) where
intrinsic lifetimes overlap signal or transit lifetimes. This occurs in the '%
\emph{spontaneous}' creation of extremely short lived virtual particles (or
'universes') due to the Heisenberg uncertainty principle. The \emph{lifetime}
of one of these particles is the '\emph{characteristic lifetime}'. An
example of this form of particle creation occurs in a '\emph{vacuum
fluctuation}' in which an electron-positron pair is produced with a lifetime
of approximately $10^{-21}$seconds (before decaying back into the \emph{%
vacuum}).

For virtual particles with energy $m_{virt}c^2$ which is less than the
uncertainty $\Delta E_{virt}$ in its total energy, the characteristic
lifetime of this \emph{virtual quantum clock }is given by:

\begin{equation}
\tau _{virt}=\Delta t_{virt}\approx \frac \hbar {\Delta E_{virt}}=\frac
\hbar {m_{virt}c^2}
\end{equation}

This is the time scale at which the characteristic lifetime and the signal
lifetime merge if you consider the emergence of the particle out of the
vacuum with velocity $v_{virt}$ and its subsequent decay back into the
vacuum as a signal along an trajectory of a finite distance, $d_{virt}$. The 
\emph{virtual signal lifetime} would then be given by:

\begin{equation}
\tau _s=\frac{d_{virt}}{v_{virt}}=\tau _{virt}
\end{equation}
This can be interpreted as a fundamental transition scale in which signals
with enough energy ($\succeq \Delta E_{virt}$) can be distinguished from a
source. It is also the energy at which information can decouple from a
source and be \emph{transmitted }as a signal.

\section{The General Theory of Time}

A general model of time is developed for complex systems composed of
networks of interacting quantum clocks. With network theory \cite{Discrete
Math} a quantum clock is treated as a vertex connected to another quantum
clock by signal paths or arcs. The network model uses the method of Feynman
Diagrams \cite{Veltman} to calculate lifetimes of states for complex systems
approaching the classical scale.

Feynman Diagrams are modified by treating the incoming and the outgoing $3$%
-momenta trajectories as \emph{arcs }(signals) between \emph{vertices }%
(clocks)\emph{\ }in a space of volume $V$. The '\emph{distance}' between the
clocks is given by the time independent metric above. The \emph{lifetime} of
the signal on the arc is given by a weight function.

The network in this case is a digraph $\mathbf{D}=(\mathbf{V},\mathbf{A})$,
where $\mathbf{V}$ is the set of vertices, and $\mathbf{A}$ is the set of
connecting arcs$.$ The \emph{signal lifetime weight function} $\tau _s:%
\mathbf{A}\longrightarrow \mathbf{\Re }$ maps the arc set $A$ to the set of
real numbers, $\mathbf{\Re }$ , giving the \emph{lifetimes} of the signals
between the vertices.

The quantum clocks represented as the vertices in the set $\mathbf{V}$ have
an \emph{intrinsic lifetime weight function} $\tau _{qc}:\mathbf{%
V\longrightarrow \Re }$ that maps \emph{intrinsic lifetimes} to the set of
real numbers. This weight function is just the inverse of $\Gamma _{qc}$ (or 
$\Gamma _{fc}$) calculated for each quantum (or Feynman) clock as calculated
above.

The separation distance $d_{i,i+1}$ is found from the \emph{metric function }%
defined by $\mathbf{d}_{metric}:\mathbf{A}\rightarrow \mathbf{\Re }$, where

\begin{equation}
\mathbf{d}_{metric}=\int_{\mathbf{r}(v_i)}^{\mathbf{r}(v_{i+1})}\mathbf{r}%
\cdot d\mathbf{r}
\end{equation}

is the path distance between vertices as travelled by a signal. Signal
trajectories may be non-Euclidean in case of the \emph{curvature }of\emph{\
space} due to massive objects. The signal lifetime is equivalent to the
intrinsic lifetime of an unstable clock whose decay is a spatial ('orbital')
transition.

The \emph{out-degree} at vertex $\nu \in \mathbf{V=}\left\{ \nu _1,\nu
_2,\cdots ,\nu _n\right\} $, is the \emph{number} of arcs (signals such as
decay products) directed away from the quantum clock. The \emph{in-degree }%
of a quantum clock is the number of arcs (signals creating or triggering a
quantum clock) directed towards the vertex (detector or scattering space).
In general these two are not equal but the sum of the $j$ \textbf{out-degree
energy-momenta, }$\mathbf{q}_i$, and the sum of the $k$ \textbf{in-degree
energy-momenta,} $\mathbf{p}_i$, are equal since total energy-momentum for
each individual quantum clock is conserved:

\begin{equation}
\left[ \left[ \sum_{i=1}^j\mathbf{q}_i\right] _{out}-\left[ \sum_{i=1}^k%
\mathbf{p}_i\right] _{in}\right] _{\nu \in \mathbf{V}}=0
\end{equation}

\subsection{Collective Excitations in Networks}

The equation for a collective or elementary excitation (phonon) lifetime of
a \emph{coupled} or \emph{continuous} set of n-quantum clocks in a network
is:

\begin{equation}
\mathbf{\tau }_{ce}\equiv \frac \hbar {\Delta E_{ce}}
\end{equation}
where $\Delta E_{ce}$ is the broadening (dispersion) in the energy levels $%
E_{ce}$ of the elementary excitations based on Heisenberg's uncertainty
principle for the phonon model (see page 345 \cite{Mattuck}). The set of
coupled clocks in a collective excitation have a non-zero potential
represented by a \emph{bond}. Collective excitations of sets of coupled
quantum clocks acting as a single unstable quantum system result in signals
(e.g. Brilloin scattering). The classical analogue is the sound wave.

When the \emph{coupling distance} associated with the \emph{bond} \cite
{cellnets} is exceeded then the collective excitation system \emph{jumps}
from a \emph{continuous} system to a \emph{discrete} system composed of a
set of decoupled quantum clocks.

\subsection{Networks of Discrete Quantum Clocks}

The lifetime of \emph{n-sequential} \emph{uncoupled} or \emph{discrete}
quantum clocks within a larger network is the $n$\emph{-sum lifetime. }For
uncoupled set of clocks the potentials between them are zero. The sequence
lifetime is given by:

\begin{equation}
\mathbf{\tau }_{nsum}\equiv \left[ \sum_{C=1}^n\left( \dfrac 1{\Gamma
_C}\right) \right] +\left[ \sum_{s=1}^n\frac{d_s}{v_s}\right] =\left[
\sum\limits_{C=1}^n\tau _C\right] +\left[ \sum\limits_{s=1}^n\tau _s\right]
\end{equation}
where $C$ is the clock number (also the network \emph{vertex} number), $%
\Gamma _C$ is the natural width of the Lorentzian energy distribution of the
final states after reconfiguration of quantum clock $C$. Between clocks $C$
and $C+1$ the signal number is $s$ (also the network \emph{arc} number), $%
d_s $ is the distance between emission and detection by a quantum clock, and 
$v_s $ is the propagation velocity of the signal along the arc. The first
term in the equation is the sum of \emph{intrinsic lifetimes }and the second
term is the sum of the\emph{\ signal lifetimes}. The net lifetime, $\tau _n$%
, from clock $C=1$ to $n$ is the lifetime of a signal propagating from clock 
$1$ to clock $n$.

If $n$ is not known then the emission of a signal appears to be created by a
single system. This signal is not the result of the decay of a collective
excitation of the system but represents the decay of the last clock in a
chain in the network.

\subsection{The Temporal Correspondence Principle}

Coupled and uncoupled sets of quantum clocks are distinguished by the
presence or absence of \textbf{bonds} respectively. The \emph{critical bond
distance}, $l_{crit}$, represents the scale at which a \textbf{temporal
phase transition} occurs between the collective and discrete representations
of quantum clocks (e.g. chain of coupled harmonic oscillators) in a network.
This assumes a single bond length for the simple case of regularly spaced
clocks in a network (e.g. crystals). For variations in clock separations
(bond lengths) collective excitations can be treated as the superposition of
spatially oriented phonons (w.r.t. the network reference frame).

Below critical separation distances, the bonds couple the clocks into a
single system with quantized energies which can be regarded as a set of
phonons. Above the critical distance, the clocks are decoupled and can act
like discrete clocks in a network. At the critical distance the two
representations overlap. This is the \textbf{temporal correspondence
principle }which states that the collective excitation and discrete n-sum
representations allow the superposition of states at the critical bond
distance.

A superposition of many different collective excitations can exist for a
system. Rotational and vibrational states can occur simultaneously. A
spectrum of collective excitations may occur if the substrate network of
coupled quantum clocks can support it.

The critical bond length represents a boundary condition (i.e. $l=l_{crit}$)
at which the two representations meet. We have the following (weak form):

\begin{equation}
\mathbf{\tau }_{ce}\Longleftrightarrow \mathbf{\tau }_{nsum}\emph{\ }
\end{equation}

where the double arrow indicates a phase transition. The two lifetimes are
not equal if the physical phase transition from a coupled state to an
uncoupled state is disjoint. In this weak case, the unstable intermediate
transition state decays in a lifetime equal to the decay lifetime of the 
\emph{last} excitation resonance that can be supported collectively by the
network before the system decouples into a discrete configuration.

In the strong form of the temporal correspondence principle the transition
is continuous at the critical bond length (at the boundary condition, the
descriptions are 'equal'). This implies:

\begin{equation}
\mathbf{\tau }_{ce}=\mathbf{\tau }_{nsum}
\end{equation}
where the lifetimes are equal. The physical meaning of the strong form is
confused by the possibility of the superposition of phonon-like excitations
in nearly discrete n-sum sets of causally connected clocks in networks. The
dispersion in the lifetimes of superimposed collective excitations supported
by the coupled network creates an uncertainty in the distinction between
coupled and uncoupled states of the system.

\subsection{Functions of 'Time'}

Virtually any \emph{time} variable calculated as a function of \emph{%
time-independent }parameters and observables can be \emph{interpreted} as
the \emph{lifetime} of a \emph{transition} between two (or more) states of
an unstable system \cite{Brout}.

For example the classical time associated with the total energy for a
two-body central force (see eqn. 3-18 \cite{Goldstein}): 
\begin{equation}
\tau _{classical\text{ }}=\dint\limits_{r_o\left( \varphi _0\right)
}^{r_f\left( \varphi _f\right) }\dfrac{dr}{\sqrt{\dfrac 2m\left(
E_I(r)-V_I(r)-\frac{l^2}{2mr^2}\right) }}
\end{equation}

can be interpreted of as an intrinsic lifetime for an 'orbital' clock
possibly in a network. The mass, $m$, maps the reconfiguration transition of
the system from an initial state $\varphi _0$ to a final state $\varphi _f$.
The 'lifetime' is evaluated at the limits of integration. The initial
position, $r_0$, is a function of the initial excited state $\varphi _0$
that 'decays' with a finite 'lifetime', $\tau _{classical}$, to the final
orbital radius, $r_f$ ,with a final state $\varphi _f$ . The 'signals'
created by this clock are the\emph{\ images} of the orbital transitions of
the mass.

\subsection{Fractal Time, Feedback, and the Messages in Signals}

\emph{Signals} generated or caused by these transitions take many forms
ranging from photons and elementary particles to nerve impulses to the
expansion of the universe. The detection of signals by systems within
networks is essential for dynamic processes such as life. \emph{Detection}
in a broad sense becomes a \emph{cause} for the evolution of complexity. The
emergent patterns of coupled and uncoupled quantum clocks in a network may
have fractal like behavior mapping \emph{fractal time }\cite{Seeley}, \cite
{Ord1}, \cite{Ord2} within emergent stochastic resonance structures \cite
{stochastic}.

It is assumed in these models that the signals in these systems do not
interact with each other. If the signals interact then \emph{signal quantum
clocks} are created by these 'collisions'. These clocks expand the network
size (i.e. number of effective clocks in the sequence) and complexity. The
treatment for interacting signals in networks is described elsewhere \cite
{Messages}.

The $n$\emph{-sum representation} emerges as the components creating
collective excitations decouple into discrete non-interacting clocks. The
decoupling may arise from increased spatial separation, increased 'action',
or changes in the physical properties and complexity of the distinct clocks.
The decoupling can lead to the emergence of increased system complexity and 
\emph{feedback} \cite{Chaos}. \emph{Feedback signals }allow system
configurations to be reset or initalized to maintain dynamical or
evolutionary processes. The lifetime associated with the feedback signal
added to the lifetime of the signal processing and generating system (e.g.
standard clocks, cells, engines, etc.) gives the net feedback cycle
lifetime. Cells are examples of systems with many superimposed levels of
feedback in which the quantum and classical models overlap in complex
chemical and metabolic networks (pathways).

These systems may be non-linear. In such cases, they appear to be cyclical
for a finite number of 'cycles'. As the behavior of the system diverges from
cyclical feedback, fractal patterns of cause and effect can emerge.
Complexity with deterministic chaos may lead to the emergence of \emph{%
fractal lifetimes} for the evolving signal pathway patterns in networks.

For a network composed of quantum clocks, the lifetime of an unstable
configuration is determined by the set of discrete (uncoupled) and
continuous (coupled) clocks acting as a single composite system. The \emph{%
messages} in $n$-body networks \cite{Messages} are carried by signals from
clock to clock. They contain the information about reconfigurations and
signal generation processes of subnetworks.

Networks with feedback form the foundation of 'memory' structures. Memory
structures process signals and generate messages containing repeatable
temporal and configuration information. \emph{Message propagators} in these
temporal networks define the signal transmission process from emission to
detection and can be mapped back to their source Feynman propagators at the
quantum clock level \cite{Messages}.

Messages can also create \emph{reconfigurations} in their targets by the
detection of the carrier signal (see section 2, axiom 5). They can encode
and create new temporal structures by modification of the detector systems
physical properties. The \emph{lifetimes of the messages} are then the \emph{%
initialization lifetimes} of the reconfiguration process of the detector.

\section{The Universe as a Quantum Clock}

An example of the 'largest' (although originally the 'smallest') quantum
clock is the Universe \cite{DeWitt}. Treating the initial state of the
universe as a quantum system with an initial wave function, $\mathbf{\Psi }%
_{U_0}$, the subsequent evolution is described by a modification of the
general features of the Schr\"{o}dinger wave equation. An example of this
modified equation is the \textbf{Wheeler-DeWitt equation (WDW)\ }\cite
{debate}, \cite{Norbury}, \cite{Gott}, \cite{Barvinsky} given by:

\[
\mathbf{H}_{wdw}\mathbf{\Psi }_{U_0}\mathbf{=0} 
\]

based on the presumed existence of a minisuperspace. There is some questions
about the validity of this assumption \cite{Asher}. If this equation is
modified for the case of \textbf{pure decay} of a non-stationary discrete
initial state coupled to a continuum of decay states then we have:

\[
\mathbf{H}_{wdw}\mathbf{\Psi }_{U_i}\mathbf{=E}_i\mathbf{\Psi }_{U_i}\mathbf{%
\neq 0} 
\]

where $\mathbf{\Psi }_{U_i}$ is the time independent wave function for the
universe, $\mathbf{H}_{wdw}$ is the Hamiltonian, and the energies, $\mathbf{E%
}_i$ , represent decay transitions. The \emph{Big Bang} can be treated as a 
\textbf{zero-order quantum clock} with a finite lifetime (e.g. for the
series of temporal and physical phase transitions such as in inflationary
scenarios). This\emph{\ initial state} represents the '\emph{first cause}'.
The initial state is \emph{not} a \emph{metastable} state. The inflationary
phase is not quantum tunneling, but a true decay process with decay products
(e.g. mass-energy structures mapped with an evolving spatial distribution
function). The\emph{\ pure decay }of the initial state of the universe can
be treated as a \emph{Feynman Clock} permitting the use of \emph{modified}
Feynman diagrams. The modification is that 'time' is omitted as a dimension
in the diagrams, but the 'lifetimes' associated with the trajectories of
signals emitted by Feynman clocks remain. \textbf{Signals} (such as the
microwave background radiation) result from the interactions between
particles and energy mediating the reconfigurations of the universe.

The '\emph{instanton}' model proposed by Hawking and Turok \cite{Hawk} is
similar to the idea of a 'pure decay' except that time is still implicitly
coupled with space. Spatialized time remains and the nature of causality is
masked by time-implicit solutions to the conventional Wheeler-DeWitt
equation.

Time independent Feynman diagrams can then be used to map solutions to the
Wheeler-DeWitt equation. Evolving structures in the early universe are
marked by temporal phase transitions (e.g. decoupling of fundamental
interactions) along with global configuration changes. The implications of
these proposals with examples of modified Feynman Diagrams will be explored
in a future paper.

\section{Conclusions}

It has been speculated that '\emph{time' }in the conventional sense\emph{\ }%
is a \emph{lifetime} associated with reconfigurations of systems and
networks of systems. By partitioning 'time' into intrinsic and signal
lifetimes, a general theory for the description of complex temporal
processes can be developed using the results of network and systems theory 
\cite{Messages}, \cite{cellnets}.

The special and general theories of time accommodate all scales of phenomena
from micro to cosmic. The thrust of this exploration has been to look at
time in a new way. The result may be a shift in a philosophical viewpoint
rather than a modification of existing paradigms. It is hoped that these
ideas will bring a new approach to looking for a solution to the 'problem of
time'.

\section{Acknowledgments}

I thank Toni, Patricia and Arleigh along with the rest of my extended
family, S. Derby, H. Kruglak, N. Boggess, E. K. Hege, G. Gilbert, J. Lubbe,
B. Buch, L. Eyres, G. Cook, E. M. High, L. Cloud, A. Balint, D. Miller, A.
Bormanis, and P. Steinhardt for their valuable discussions and support. I
also would like thank the staff at the National Superconducting Cyclotron
Laboratory for the space to think about time.

\end{document}